\documentclass[prl,a4paper,twoc olumn]{revtex4} 

\usepackage[draft]{hyperref}
\usepackage{amsmath}
\usepackage{graphicx}
\usepackage{epstopdf}
\usepackage{comment}
\usepackage{soul}
\usepackage{color}
\usepackage[dvipsnames]{xcolor}

\begin{document}

\title{Improving Polarisation Squeezing In Sagnac Interferometer Configuration Using Photonic Crystal Fibre}


\author{Morgan J. Tacey,$^{1,*}$ Joel F. Corney,$^1$}

\address{
$^1$School of Mathematics and Physics, The University of Queensland, \\  Brisbane, QLD, 4072, Australia
\\
$^*$Corresponding author: tacey@physics.uq.edu.au
}

\begin{abstract}
The greater confinement of light that is possible in photonic crystal fibres leads to a greater effective nonlinearity, which promises to yield greater quantum squeezing than is possible in standard optical fibre.   However, experimental work to date has not achieved improvements over standard fibre. We present a comprehensive numerical investigation of polarisation squeezing in photonic crystal fibre in a Sagnac configuration.  By including loss, a non-instantaneous Raman response, excess phase-noise, second- and third-order dispersion and self-steepening, the simulations are able to identify the physical factors that limit current photonic crystal fibre squeezing experiments.
\end{abstract}


\maketitle

\noindent Squeezed light is an important resource for the fields of quantum information \cite{NeergaardNielsen:2010by,Braunstein:2005bp}, where it is a route to quantum entanglement \cite{Bowen:2002cj,Korolkova:2002hpa,Wagner:2008jw}, and precision measurement, where for example it is being used to increase the precision of gravity wave detectors \cite{Collaboration:2011cb}.
It can be generated simply by an intense light pulse propagating in an optical fibre \cite{Drummond:1993va}, via the optical Kerr effect \cite{STOLEN:1973p489} which is induced by the intrinsic third-order or $\chi^{(3)}$ nonlinearity.

The advantages of fibre based squeezing include its flexibility of wavelength and power, avoidance of the phase-matching required in $\chi^{(2)}$ materials, low intrinsic losses, relatively low cost, long form factors and ability to integrate with existing optical network infrastructure.  It also provides a source of continuous variable entanglement \cite{Braunstein:2005bp}, useful in unconditional quantum information experiments.  The biggest disadvantage is two dominant noise sources that inhibit the degree of squeezing. Raman noise scales inversely with pulse width and guided acoustic wave Brillouin scattering (GAWBS) scales with fibre length. The pursuit of better squeezing in optical fibre becomes an exercise in controlling these noise sources.

Polarisation squeezing has to date achieved the best result in optical fibre, with 6.8 dB \cite{Dong:2008bx} below shot noise. One promising solution to curtail the noise sources is photonic crystal fibres (PCF). The higher confinement in PCF means that the same Kerr effect can be achieved with shorter interaction lengths or lower pulse intensity than standard optical fibre. A shorter fibre length means a shorter distance for classical GAWBS noise to accumulate.  On the other hand, a less intense pulse, with a broader pulse shape, will not excite the Raman response to the same extent.   Despite this potential, improved squeezing with PCF was been difficult to achieve experimentally \cite{Milanovic:2007hz}, with 3.9 dB below shot noise the best achieved to date \cite{Milanovic:2010ij}.

In this paper we present the results of comprehensive numerical simulations of the PCF squeezing experiments of Milanovic {\rm et al} \cite{Milanovic:2010ij}.  The physical factors that prevented the PCF experiments from achieving the same levels of squeezing as those achieved with standard fibre are identified to be an excess classical phase noise and an additional effective loss arising from reduced fringe visibility.

The experiment employed a Sagnac configuration of counter-propagating pairs of pulses, which was found to be an improvement over the earlier single-pass polarisation squeezing  \cite{Milanovic:2007hz}.
  After a 50:50 splitting, ultrashort soliton pulses counter-propagate through a 1m length PCF in the same polarisation plane. The squeezed pulses are then combined after a $\pi/2$ relative phase shift to form a circularly polarised pulse, whose fluctuations in the dark Stokes $\hat{S}_2$-$\hat{S}_1$ plane are then measured by use of a polarisation rotator and a pair of balanced photo detectors. Here we define the Stokes operators as $\hat{S}_0 = \hat{N}_{xx} + \hat{N}_{yy}$, $\hat{S}_1 = \hat{N}_{xx} - \hat{N}_{yy}$, $\hat{S}_2 = \hat{N}_{xy} + \hat{N}_{yx}$, and $\hat{S}_3 = i\hat{N}_{yx} - i\hat{N}_{xy}$ where $\hat{N}_{ij} = \int dz \hat{\Psi}_i^\dagger (t,z) \hat{\Psi}_j (t,z)$ and $\hat{\Psi}_i^\dagger, \hat{\Psi}_i$ are the quantum field operators for horizontal and vertical polarisations, $i=x,y$ respectively. 
A half-wave plate is used to access different polarisation measurements by rotating the Stokes operators as $\hat{S}_\theta = \cos{(\theta)}\hat{S}_1 + \sin{(\theta)}\hat{S}_2$ in the dark plane \cite{Dong:2008bx,Corney:2008p2}.

For polarisation measurements, the Heisenberg inequality depends on the mean polarisation: $\Delta^2 \hat{S_i}\Delta^2 \hat{S_j} \geq \bigl| \bigl< \hat{S_k}\bigr>\bigr|^2$, where the variance is $\Delta^2 \hat{S_i} = \bigl<\hat{S_i}^2\bigr> - \bigl<\hat{S_i}\bigr>^2$. Thus squeezing in the dark plane occurs when $\Delta^2 \hat{S_\theta} < | \langle \hat{S_3}\rangle |$ \cite{Dong:2008bx,Corney:2008p2}, resulting in the polarisation squeezing metric $M = {\Delta^2 \hat{S_{\theta}}} / {\bigl| \bigl< \hat{S_3} \bigr> \bigr| }$. 
After linear loss $\epsilon$, the squeezing metric becomes $M = (1/\langle\hat{S}_0\rangle) [(1-\epsilon)\langle\hat{S}_1\rangle + \epsilon \langle\hat{S}_0\rangle ]$.

The simulations are performed by numerically solving a stochastic nonlinear Schrodinger equation derived in the truncated Wigner representation \cite{Drummond:1993p279}, which has proven successful in polarisation squeezing simulations for standard fibre \cite{Corney:2008p2,Dong:2008bx,Corney:2011wv}.  Scaling time and distance by the pulse width $t_0$ and dispersion length  $z_0=t_0^2/\beta_2$, respectively, and using a propagating frame of reference at the group velocity $v$, we can write the full evolution equation as:

\begin{align}
\label{eqn:PDEModel}
\frac{\partial \phi(z,t)}{\partial z} &= i \left[ \frac{1}{2}\frac{\partial^2}{\partial t^2}  -i B_3\frac{1}{6}\frac{\partial^3}{\partial t^3}  + G\eta \right. \nonumber \\
&\quad \left. +\int_{-\infty}^{\infty}dt' h(t-t')|\phi(z,t)|^2 + \Gamma(z,t) \right] \phi(z,t)\nonumber \\ 
&\quad - s \frac{\partial |\phi(z,t)|^2\phi(z,t)}{\partial t},
\end{align}
where we have included third-order dispersion with relative strength, $B_3 = \beta_3/|\beta_2| t_0$ and  self-steepening  with $s = 1/\omega_0 t_0$.  The nonlinear response $h(\tau)$ includes the Raman contribution, which has an associated noise term $\Gamma$. (See \cite{Drummond:2001p1} for details.)  GAWBS is modelled as Gaussian fluctuations  $\eta$ in the refractive index, with a magnitude $G=3.2\times10^{-4}$  that was determined by a single-parameter fit to the low-pulse-energy squeezing experimental results \cite{Milanovic:2010ij}.

The field is rescaled as $\phi = (v t_0 /\bar{n})\Psi$ where $\bar{n}=|\beta_2| /(\hbar \omega \gamma t_0)$  is half the average photon number for the fundamental soliton at frequency $\omega_0$ in a medium with nonlinear parameter $\gamma$. We base the fibre parameters on a NKT Photonics NL-PM-750 nonlinear PCF \cite{JW:2010wr} at a wavelength of 810 nm.  The effective modal area of the PCF, $2.0\times10^{-12}$ m$^2$, is much smaller than the usual optical fibre, leading to a larger effective nonlinearity $\gamma=91.4$ (kmW$)^{-1}$.  With group velocity dispersion $\beta_2=12.2$ps$^2/$ km and pulse width $t_0 =  0.068$ ps, the dispersion length is $z_0 = 0.38$ m and characteristic fundamental photon number $\bar{n}=8.0\times10^6$.  The energy of a fundamental soliton is then $E_s = 2\overline n \hbar \omega = 3.92$ pJ.


\begin{figure}[htp]
\centering
\includegraphics[width = 0.49\columnwidth]{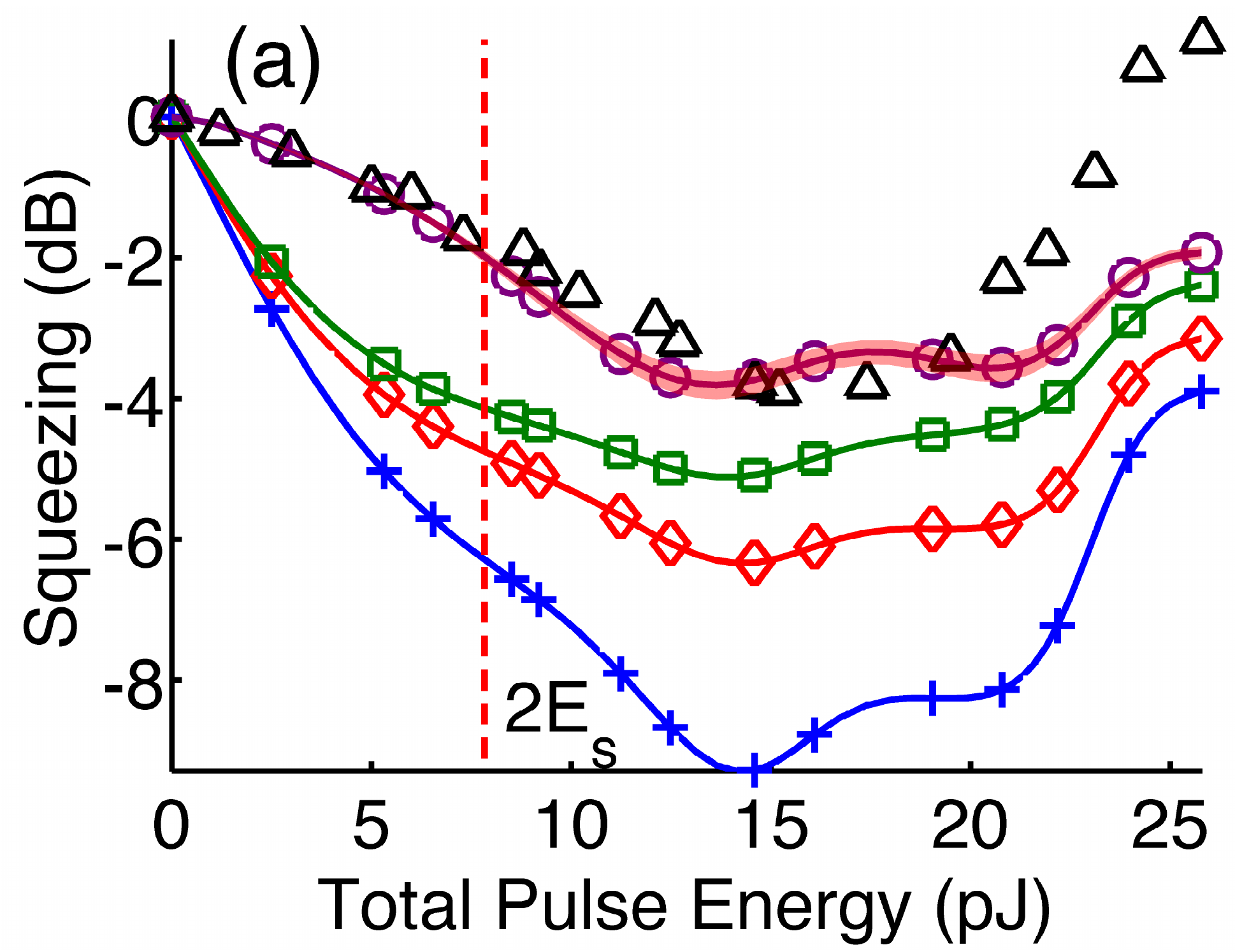}
\includegraphics[width = 0.49\columnwidth]{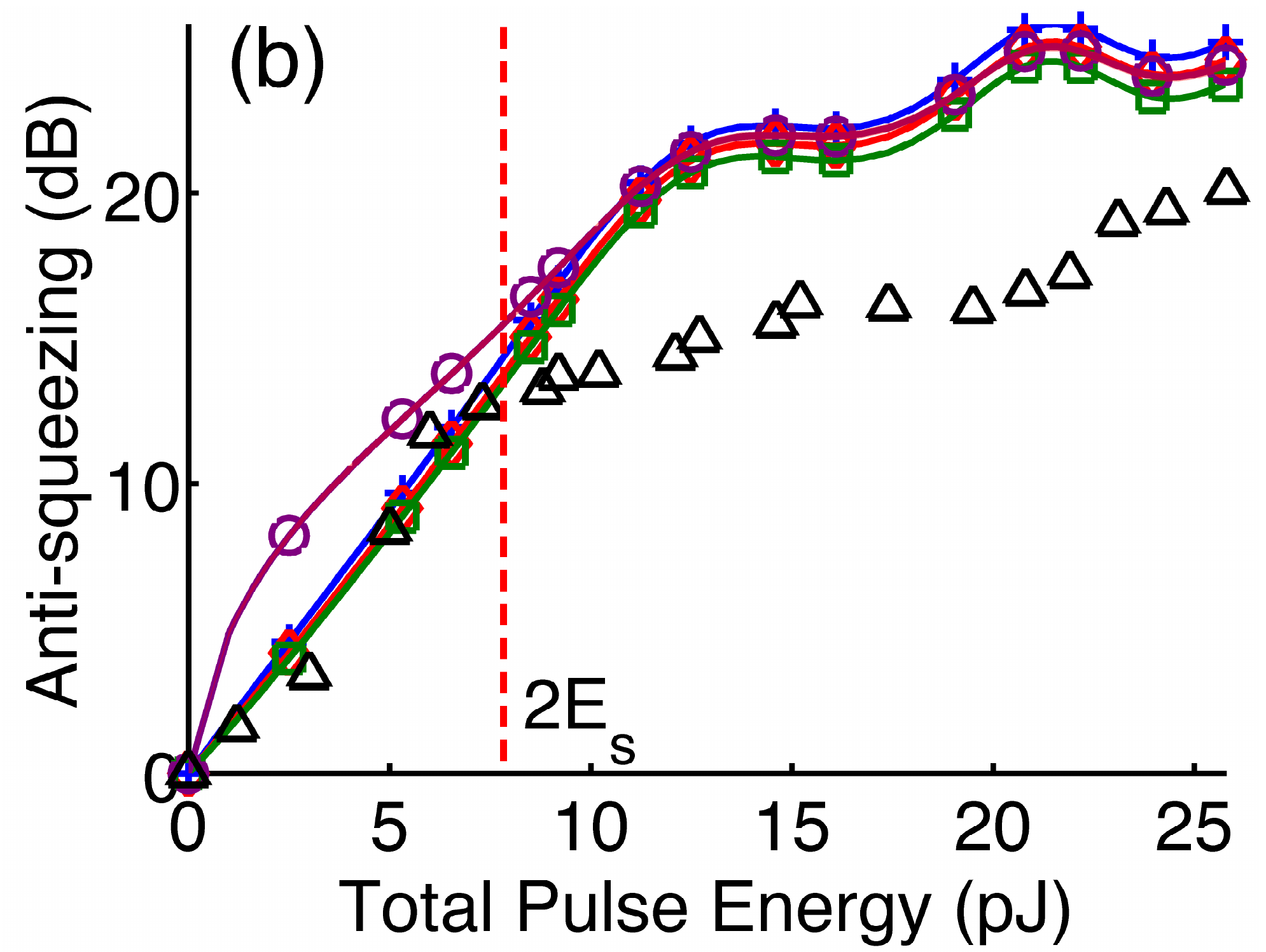}
\includegraphics[width = 0.49\columnwidth]{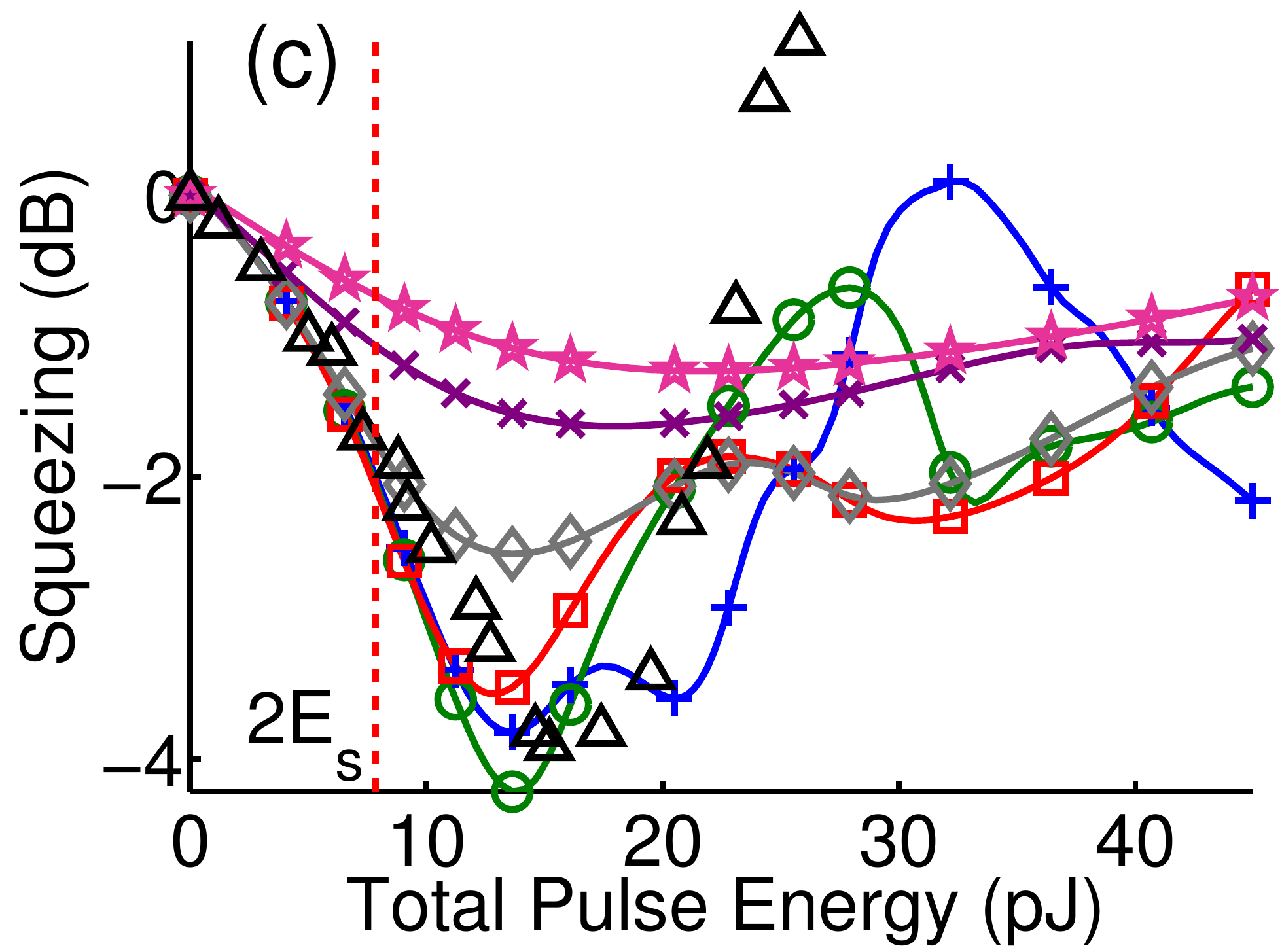}
\includegraphics[width = 0.49\columnwidth]{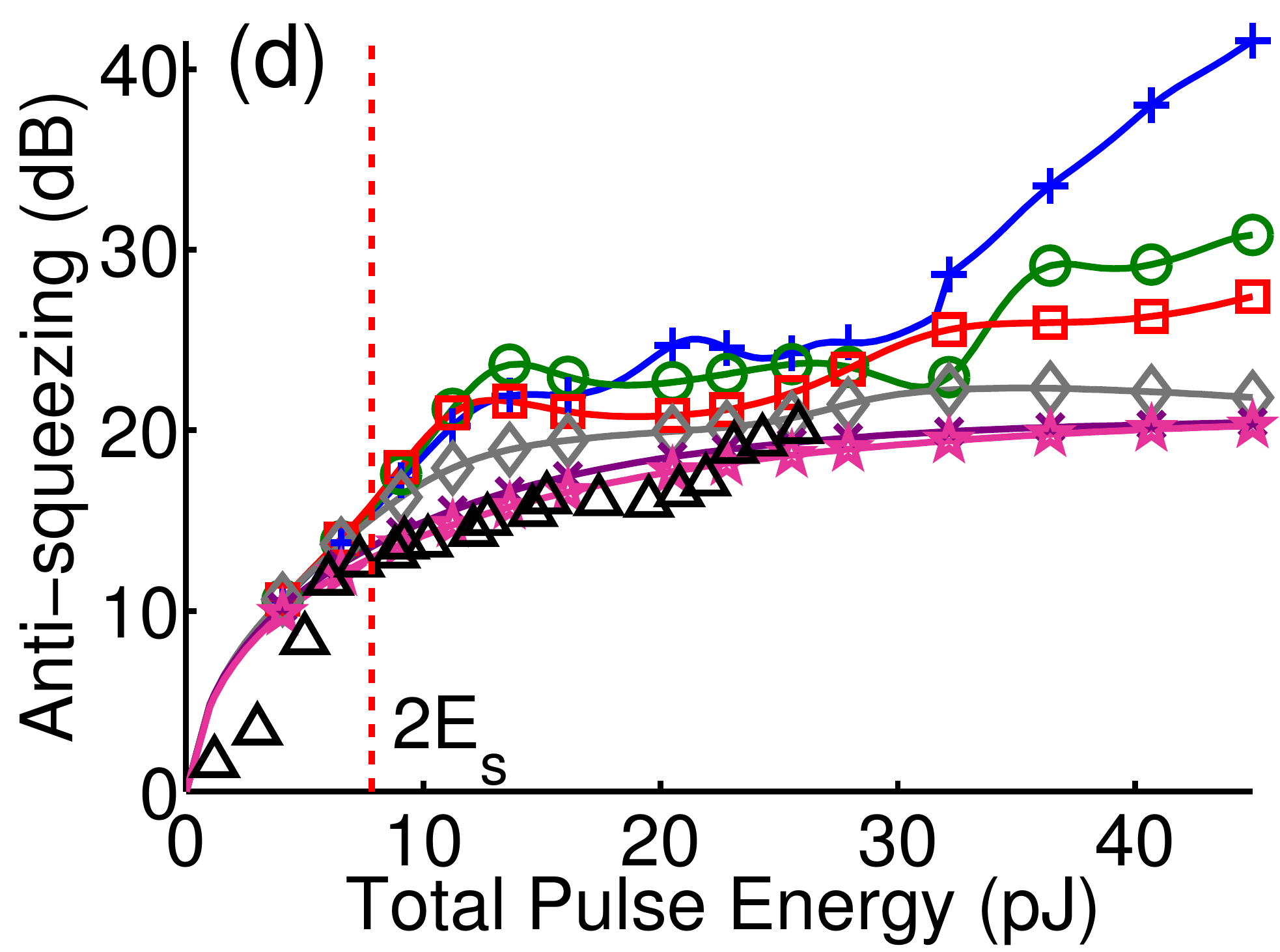}
\caption{(a) Squeezing and (b) anti-squeezing as a function of the total pulse energy.  Pluses show the `pure' squeezing, with no higher-order dispersive effects ($B_3=0$, $s=0$), GAWBS noise (G = 0) or loss.  The remaining curves successively add in the effects of the 13\% linear loss (diamonds),  imperfect spectral overlap (squares) and GAWBS noise (circles).  The shading on the latter indicates the effect of a $\pm0.03$ uncertainty in the experimentally determined spectral overlap  \cite{Milanovic:2010ij}.  Experimentally measured squeezing are given by the triangles. Vertical dashed line indicates the energy of two soliton inputs. (c) Squeezing and (d) anti-squeezing for various amounts of third-order dispersion $B_3$:  0 (pluses), 0.3 (circles), 0.5 (squares), 1.0 (diamonds), 3.0 (crosses) and 6.0 (stars). Other parameters are as in Fig. 1a-b, with experimental results again plotted as triangles. $10^4$ stochastic paths were used in the simulation to give a maximum sampling error of  $\pm6.3\times10^{-2}$ dB.}
\label{fig1}
\end{figure}

A comparison of the experimental and numerical squeezing results, over the range of pulse energies used in the experiment, is given in Figure \ref{fig1}a, \ref{fig1}b.  Even without taking into account any linear losses or GAWBS, the theoretical squeezing results (crosses) are consistent with the experimental data in the overall trends and in the energy at which best squeezing occurs (14.6 pJ). In general, a larger pulse energy leads to a stronger Kerr effect, until Raman effects combined with pulse reshaping cause the squeezing to deteriorate.  Quantitative agreement is obtained once we take into account linear losses (diamonds), the finite overlap \cite{Milanovic:2010ij} of the pulses (squares), and GAWBS phase noise (circles). The best squeezing found numerically was -5.1 dB without phase noise, and -3.8 dB with phase noise, consistent with the measured -3.9$\pm$0.3 dB.

Significant differences between the simulations and experiment start to emerge once the pulse energy exceeds about twice that needed to generate solitons.  In this regime, where soliton fission and pulse reshaping lead to complex pulse evolution, we can expect that small discrepancies between simulation and experiment can lead to large differences in the squeezing.  Moreover, higher-order effects may become important, due to pulse narrowing.  Here we considered two such effects: self steepening and third-order dispersion. We found the results with self-steepening are almost indistinguishable from those without. The maximum difference of 0.08 dB is of the same order of magnitude as the maximum sampling error of $\pm6.3\times 10^{-2}$ dB in the simulation.

Since the size of the third-order dispersion is unknown, we plot in Fig.\ \ref{fig1}c and \ref{fig1}d results for a range of possible values, $B_3=[0,6]$.  As expected, it has a stronger effect at higher pulse energies, and a value of $B_3\simeq 0.3$ may explain some of the faster deterioration of squeezing observed experimentally.

\begin{figure}[htp]
\centering
\includegraphics[width = \columnwidth]{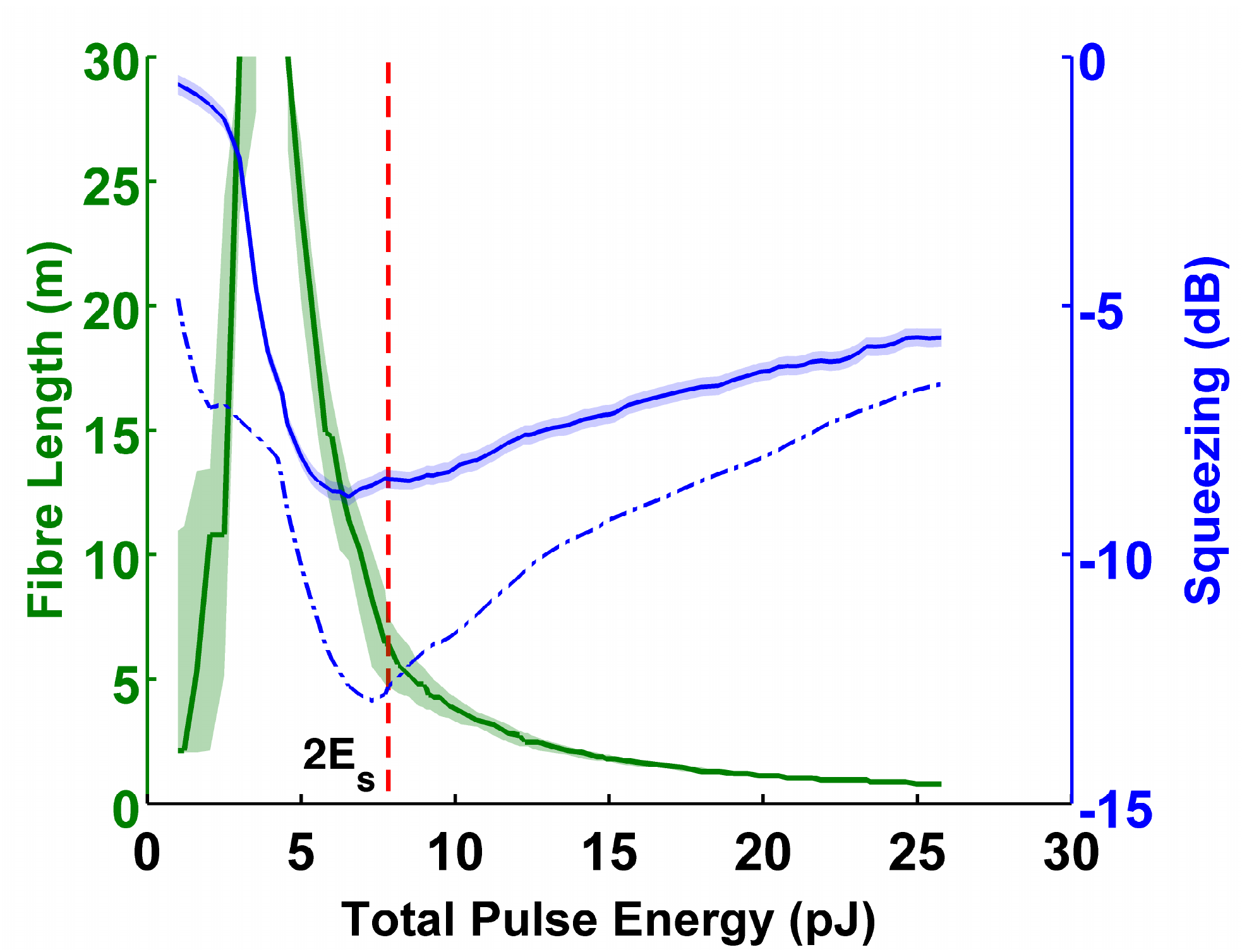}
\caption{Fibre length (shaded green line, left axis) for optimal squeezing with GAWBS noise (shaded blue line, right axis, with maximum sampling error of $\pm0.2$ dB) as a function of total pulse energy, with (solid line) and without (dashed-dot line) GAWBS noise. The green shading shows the range of fibre lengths that give squeezing within 5\% of the optimum. For some of the small pulse energies (between 3.54 pJ and 4.57 pJ) the trend indicates marginally better squeezing for fibre lengths larger than 30 m. Linear loss and finite spectral overlap effects have not been included. $10^3$ stochastic paths were used in the simulation.}
\label{fig2}
\end{figure}

So far, we have only considered the squeezing of a fibre with a fixed length of $L = 1.0$ m.  Figure \ref{fig2} shows the squeezing optimized over fibre length (up to 30 m).  Although the dependence of fibre length is relatively weak, as indicated by the breadth of the shaded area, significantly better squeezing can be obtained by use of longer fibres.  Overall the best squeezing was found to be -12.6 dB (without any linear losses or GAWBS noise) using soliton inputs and a fibre length of around 5m, compared with -9 dB achieved with 1m of fibre and higher input energies.

We conclude by noting that, from amongst the various physical effects that we have considered, there are two that are particularly pertinent for the experimental setup considered here: GAWBS noise and the imperfect spectral overlap of the counter-propagating  pulses.   Here we see GAWBS reducing the optimal squeezing from -12.6 dB to -8.3 dB (without linear loss), wheres for the previous single-pass scheme using conventional fibre \cite{Dong:2008bx}, GAWBS noise had a negligible effect on the optimal squeezing.  The imperfect spectral overlap introduces an extra effective linear loss that was not present in the previous work.  However, if these effects can be eliminated in future experiments, we can expect to see a gain in the degree of squeezing achieved by use of PCFs, over and above what has been observed in standard fibre.  One potential benefit of using PCFs that is yet to be explored is the scope for reducing detrimental Raman effects, through use of pulses with larger time width.

\bigskip



 \bibliographystyle{osajnl}
 \bibliography{jfc_papers,m_papers}

\end{document}